# Spin-orbit coupling induced semi-metallic state in the 1/3 hole doped hyper-kagome $Na_3Ir_3O_8$

Tomohiro Takayama[1,*], Akiyo Matsumoto[2], Jürgen Nuss[1], Alexander Yaresko[1], Kenji Ishii[3], Masahiro Yoshida[3,4], Junichiro Mizuki[3,4] & Hidenori Takagi[1,2]

[1]Max Planck Institute for Solid State Research, Heisenbergstrasse 1, D-70569 Stuttgart, Germany
[2]Department of Physics and Department of Advanced Materials, University of Tokyo, 7-3-1 Hongo, Bunkyo-ku, Tokyo 113-0033, Japan
[3]SPring-8, Japan Atomic Energy Agency, Sayo, Hyogo 679-5148, Japan
[4]School of Science and Technology, Kwansei Gakuin University, Sanda, Hyogo 669-1337, Japan

The complex iridium oxide $Na_3Ir_3O_8$ with a B-site ordered spinel structure was synthesized in single crystalline form, where the chiral hyper-kagome lattice of Ir atoms, as observed in the spin-liquid candidate $Na_4Ir_3O_8$, was identified. The average valence of Ir is 4.33+ and, therefore, $Na_3Ir_3O_8$ can be viewed as a doped analogue of the hyper-kagome spin liquid with $Ir^{4+}$. The transport measurements showed that $Na_3Ir_3O_8$ is in fact a semi-metal. The electronic structure calculation demonstrated that the strong spin-orbit coupling of Ir yields the semi-metallic state out of an otherwise band insulating state, which may harbor exotic topological effects embedded in the hyper-kagome lattice.

The realization of a quantum spin liquid is a long-sought dream in condensed matter physics, where exotic phenomena like fractional excitations or unconventional superconductivity nearby are anticipated[1]. The most promising arena for a spin liquid is a geometrically frustrated lattice based on a triangular motif. Antiferromagnetically interacting spins on such a frustrated lattice cannot simultaneously satisfy all magnetic bonds and, as a result, macroscopic degeneracy could remain in the ground state. To date the experimental efforts have put forward several candidates such as transition metal oxides with a kagome lattice[1,2] and organic Mott insulators with a triangular lattice[1,3].

The recent discoveries of candidate materials include $Na_4Ir_3O_8$ which emerged as the first candidate for a three-dimensional quantum spin liquid[4]. In this complex oxide, Ir atoms with a localized $S = 1/2$ moment (or more likely close to $J_{eff} = 1/2$ moment, see below) form a corner-sharing network of triangles in three-dimensions, called "hyper-kagome" lattice. All the Ir sites and the Ir-Ir bonds are equivalent, rendering the hyper-kagome lattice magnetically frustrated. Indeed, $Na_4Ir_3O_8$ exhibited no magnetic ordering down to 2 K despite the strong antiferromagnetic interaction inferred from the Curie-Weiss temperature $\theta_W$ of -650 K. This discovery triggered intensive experimental and theoretical surveys on this compound, including proposals for the presence of a spin Fermi surface[5].

In tandem with the discovery of $Na_4Ir_3O_8$, complex $Ir^{4+}$ oxides have recently emerged as a novel playground for physics of strong spin-orbit coupling (SOC). The SOC of $Ir^{4+}$ is as large as $\lambda_{SO} \sim 0.6$ eV, reflecting the heavy atomic mass. $\lambda_{SO}$ is not small at all as compared with the other parameters dominating the electronic states, such as the inter-site hopping $t$, the Coulomb repulsion $U$ and the crystal field splitting $\Delta$. The interplay of the large SOC with the other parameters leads to the formation of unprecedented electronic phases. In the layered perovskite $Sr_2IrO_4$, for example, the spin and orbital degrees of freedom are intimately entangled to produce $J_{eff} = 1/2$ states, giving rise to a novel spin-orbital Mott insulator[6]. The magnetic coupling of $J_{eff} = 1/2$ moments in such spin-orbital Mott insulators can be distinct from those of the spin-dominant moments in $3d$ oxides, as exemplified by the possible Kitaev spin-liquid proposed for honeycomb iridates[7]. Such unique magnetic couplings in iridates would make the mystery of the possible spin-liquid state of $Na_4Ir_3O_8$ even more intriguing[8].

The insulating state of $Na_4Ir_3O_8$ is marginally stabilized by a modest $U$ with the help of strong SOC. Such weak Mottness, implying the close proximity to a metallic state, has been discussed to play a vital role in realizing the spin-liquid ground state[9], along with organic triangular spin liquids[1]. It should not be very difficult to bring $Na_4Ir_3O_8$ to a metallic state by carrier-doping or by applying pressure. Exotic superconductivity at the critical boarder to a spin liquid might be anticipated in analogy with the organic systems[10] and is worthy for exploration.

Despite such intriguing outlooks in $Na_4Ir_3O_8$, not much progress has been achieved in the critical investigation of the spin-liquid like state including the role of spin-orbit coupling, largely due to the lack of single crystals. During the course of attempting to grow single crystals of $Na_4Ir_3O_8$, we obtained single crystals of $Na_3Ir_3O_8$, a B-site ordered spinel. The crystal structure is distinct from that of $Na_4Ir_3O_8$, a spin liquid candidate, but shares the same Ir-O hyper-kagome network. $Na_3Ir_3O_8$ therefore can be viewed as a doped hyper-kagome spin liquid. We report here that $Na_3Ir_3O_8$ has a semi-metallic ground state produced by the strong spin-orbit coupling.

**Results**

**Crystal Structure Analysis**. X-ray diffraction analysis and detailed refinement of the crystal structure was performed on $Na_3Ir_3O_8$ single crystals grown by a flux method (see Methods). Satisfactory refinement was obtained with space groups $P4_132$ or $P4_332$. Accordingly, the crystal under investigation turned out to be a racemic twin (see Supplementary). The refined structural parameters are listed in Table 1. The structure shown in Fig.1**a** can be viewed as an ordered spinel, an intimately related but distinct structure to that of polycrystalline $Na_4Ir_3O_8$. Rewriting the chemical formula of 1/2 $Na_3Ir_3O_8$ as $Na(Na_{1/4}, Ir_{3/4})_2O_4$, in correspondence with that of spinel $AB_2O_4$, is convenient to understand the structure. Na(2) in Table 1 corresponds to the tetrahedral A-site of spinel structure. In $Na_4Ir_3O_8$, this site is empty and, instead, the octahedral A-sites are partially occupied by Na. Na(1) corresponds to 1/4 of the pyrochlore B sub-lattice. The remaining 3/4 of the pyrochlore B sub-lattice sites are occupied by Ir. All the Ir sites are equivalent and form the same hyper-kagome network as in $Na_4Ir_3O_8$. The presence of chirality in the hyper-kagome network gives rise to two possible space groups with different chiralities as illustrated in Fig. 1**b**. We did not find any signature of non-stoichiometry in the refinement and therefore the composition of single crystal should be $Na_3Ir_3O_8$. We note that the structures of $Na_3Ir_3O_8$ and $Na_4Ir_3O_8$ cannot be continuously converted with each other due to the different Na positions and thus they are two distinct crystallographic phases.

**Transport and Magnetic Properties.** The composition of $Na_3Ir_3O_8$ corresponds to Ir valence of 4.33+, not 4+ i.e. $5d^5$ configuration. Considering that $Na_4Ir_3O_8$ is an $Ir^{4+}$ ($5d^5$) Mott insulator, $Na_3Ir_3O_8$ may be viewed as 1/3 hole doped Mott insulator, namely 1/3 hole doped hyper-kagome spin liquid. The obtained single crystals were indeed found to show metallic behavior of resistivity as shown in Fig. 2**a**, in marked contrast to the spin liquid $Na_4Ir_3O_8$. The magnitude of resistivity was relatively large as a metal, ~1 mΩcm at 5 K. The Hall coefficient of $Na_3Ir_3O_8$ indicated that the poorly metallic behavior originates from the low carrier concentration. The magnitude of Hall coefficient (Fig. 2**b**) was significantly large as a metal and negative, indicative of the low carrier concentration and the dominant electron-like contribution to the transport. The Hall mobility at 5 K is as large as ~100 $cm^2$/V·s and, clearly, the disorder is not the dominant factor of the poorly metallic behavior. The carrier number estimated from the Hall constant is of the order of $10^{19}$ $cm^{-3}$ at 5 K, which is too small to be accounted as a simple 1/3 hole-doped Mott insulator with 2/3 electrons. The rapid increase of Hall coefficient with temperature despite the metallic behavior of resistivity, more than one order of magnitude from 5 K to 300 K, indicates the coexistence of two different types of carriers. This implies that $Na_3Ir_3O_8$ is either a semi-metal or a very lightly doped narrow gap semiconductor. Unexpectedly, the ground state of a doped hyper-kagome appears to be very close to a band insulator.

The low temperature specific heat of $Na_3Ir_3O_8$, shown in the inset of Fig. 2**a**, yields an electronic specific heat coefficient $\gamma = 4.3$ mJ/Ir-mol·K$^2$. This value is pronounced considering the vanishingly small density of carriers and therefore implies the presence of heavy mass carriers, namely flat band. The magnetic susceptibility (Fig. 2**c**) is almost temperature-independent, distinct from that of magnetic $Na_4Ir_3O_8$. The paramagnetic susceptibility was estimated to be $\chi_0 = 2.7 \times 10^{-4}$ emu/Ir-mol by subtracting the contribution of core diamagnetism taken from the values for $Na^+$, $Ir^{4+}$ and $O^{2-}$. This value gives the Wilson ratio ~ 4.6 combined with $\gamma = 4.3$ mJ/Ir-mol·K$^2$, apparently beyond the value of correlated electron regime. We speculate the enhanced $\chi_0$ in part originates from Van Vleck–like process.

**Resonant Inelastic X-ray Scattering.** To verify the 1/3-doped state, we performed a resonant inelastic X-ray scattering (RIXS) on $Na_3Ir_3O_8$ single crystals and a $Na_4Ir_3O_8$ polycrystalline sample as a reference. $L_3$-edge excitation of Ir was used where a $2p_{3/2}$ core electron is excited into $5d$ orbitals and then de-excited[11]. In the obtained RIXS spectra shown in Fig. 3, three peaks at around 0.2, 1.0 and 4.0 eV are clearly observed. All those peaks showed only very small dispersions, which very likely originates from the intra-atomic excitations within the $d$-orbitals of Ir. The intra-atomic character of excitations justifies the reasonable comparison of single crystal $Na_3Ir_3O_8$ and polycrystalline $Na_4Ir_3O_8$. We could see the clear shifts of the peaks at 1.0 and 4.0 eV to lower energy in $Na_4Ir_3O_8$. It appears that this is due to the shift of Fermi level by the 1/3 hole doping but the suppressed energy scale of crystal field effect in $Na_4Ir_3O_8$, due to the increased Ir-O distance (~3%), could be in fact the dominant cause of the shift. The absence or suppression of broad structure around 0.2 eV in $Na_4Ir_3O_8$, in contrast, cannot be accounted for simply by the crystal field effect and very likely reflects the difference of band filling between the two compounds, which will be justified based on the result of band calculation described below.

**Electronic Structure Calculation.** *Ab-initio* electronic structure calculation using a fully relativistic LMTO code[12] revealed that $Na_3Ir_3O_8$ is a compensated semi-metal due to the interplay of periodic potential and SOC, which is consistent with the experimental observation described above. Figure 4 depicts the electronic state around the Fermi energy where $t_{2g}$ orbitals of Ir have a dominant contribution. In iridates, the $5d$ electrons are accommodated into the $t_{2g}$ manifolds due to large $t_{2g}$ – $e_g$ crystalline field splitting. In spite of the non-integer Ir $5d^{4.67}$ filling with 14 $t_{2g}$ electrons per 3 Ir in a formula unit, a gap of 0.2 eV opens within $t_{2g}$ bands in the scalar-relativistic calculation neglecting SOC as shown in Fig. 4**a**. A similar gap separating $t_{2g}$ bands was also found for $Na_4Ir_3O_8$ and explained by strong *p-d* hopping[13]. In this case, due to higher Na content the $t_{2g}$ bands above the gap are partially filled.

The strong SOC of Ir, in reality, splits the conduction and the valence bands substantially and a negative band gap is enforced. Bands calculated by including SOC are shown in Fig. 4**c**. Since the crystal structure lacks inversion symmetry, SOC lifts the Kramers degeneracy everywhere except for time-reversal invariant points. If the SOC strength is gradually tuned from 0 to its calculated value a pair of unoccupied $t_{2g}$ bands, colored in magenta, starts to bend down near the R point, closes the gap, and, finally, crosses six-fold degenerate bands lying at -0.1 eV at the R point. This pair of bands creates two electron-like Fermi surfaces around the R point. In order to maintain the charge balance, hole-like bands (red) near the Γ point become partially depopulated. As a result, $Na_3Ir_3O_8$ is a semi-metal with electron pockets around the R point and hole pockets around the Γ point, which seems to be protected by the degeneracy at the R point below $E_F$. The calculation indicated enhanced mass carriers; 1.4, 2.3, 3.7 and $5.6m_0$ for hole bands and two electron bands with $1.8m_0$, presumably reflecting the narrow width of SOC reconstructed bands. Those masses yield $\gamma_{calc}$ = 2.9 mJ/Ir-mol·K$^2$, which is close to and only 35% smaller than the experimentally observed $\gamma$ = 4.3 mJ/Ir-mol·K$^2$. The more dispersive electron bands with higher mobility result in negative calculated $R_H$ which is consistent with the observed sign of the Hall constant. The recent observation of Fano resonances in the phonon spectra of $Na_3Ir_3O_8$ is fully consistent with the semi-metallic electronic state where the electron-hole excitation continuum interferes with superimposed discrete phonon states[14].

**Discussion**

The most striking effect of SOC in $Na_3Ir_3O_8$ is the closure of scalar relativistic gap. This is in stark contrast to the case of $SrIrO_3$, where the LDA calculation without SOC yields a metal rather than a band insulator. SOC reconstructs metallic $t_{2g}$ bands and leads to the semi-metallic state due to the band split[15]. An inspection of orbitally resolved densities of Ir $5d$ states reveals that only two of three $t_{2g}$ states of each Ir ion contribute to the unoccupied $t_{2g}$ bands whereas the third $t_{2g}$ orbital is completely filled, in the scalar-relativistic calculation shown in Fig. 4**b**. The energy of the filled orbital is lower because the average Ir-O distance in the plane containing it is somewhat larger than in the planes of the other two orbitals. However, the 14-to-4 ratio of the number of occupied and unoccupied $t_{2g}$ bands implies that the gap cannot be simply caused by local distortions of $IrO_6$ octahedra. We may understand the gap formation by considering $Ir_3$ triangular molecules which are the basic structural unit of the hyper-kagome network. Each $Ir_3$ molecule in $Na_3Ir_3O_8$ accommodates an even number of $t_{2g}$ electrons when 1/3 hole is doped to $5d^5$ $Ir^{4+}$, 14 electrons for 18 quasi-molecular orbitals, which could make the system a band insulator.

SOC splits five-fold degenerate $d$-orbitals into $j = 5/2$ and $3/2$ characters. In an isolated octahedron, SOC split states are mixed due to the crystalline field, and form a lower lying quartet with $j = 5/2$ and $3/2$ admixture ($J_{eff} = 3/2$ state), a doublet with pure $j = 5/2$ character ($J_{eff} = 1/2$ state), and upper doubly-degenerate $e_g$ orbitals[16]. Ir $5d^{4.67}$ configuration indicates that $E_F$ locates within the $J_{eff} = 1/2$ manifolds, which can be expressed as the equal superpositions of three $t_{2g}$ orbitals. Such $J_{eff} = 1/2$ picture was also reported for $Na_4Ir_3O_8$[13]. An analysis of the orbital character of the $t_{2g}$ bands for $Na_3Ir_3O_8$, depicted in Fig. 4**d**, shows that the bands below -0.5 eV have predominantly $J_{eff} = 3/2$ character. The bands just below the $E_F$ have an almost pure $J_{eff} = 1/2$ character, indicating that SOC entangles the three $t_{2g}$ orbitals in an equal weight unlike the scalar relativistic calculation described above. On the other hand, the unoccupied bands have significant admixture of $J_{eff} = 3/2$ states. This implies rather strong hopping between $J_{eff} = 1/2$ and $J_{eff} = 3/2$ states, which is likely inherent to the edge-sharing network of $IrO_6$ octahedra[17,18].

Based on the calculated electronic structure, we can reasonably assign the relevant $d$-$d$ excitations to the peaks observed in the RIXS spectrum in Fig. 3. The peak structures observed around 1.0 and 4.0 eV correspond to the excitations from the $J_{eff} = 3/2$ to the unoccupied $t_{2g}$ bands and from the $t_{2g}$ to $e_g$ bands, respectively. The peak around 0.2 eV originates from the transition between the occupied $J_{eff} = 1/2$ bands and the unoccupied $J_{eff} = 1/2$ bands with $J_{eff} = 3/2$ admixture, likely representing the strong inter-band transition between the pair of flat bands crossing $E_F$. The absence of the 0.2 eV peak in the RIXS spectrum of $Na_4Ir_3O_8$ is fully consistent with the filling of the unoccupied and flat $J_{eff} = 1/2$ bands associated with the removal of 1/3 hole in $Na_4Ir_3O_8$.

In summary, we successfully synthesized single crystals of hyper-kagome iridate $Na_3Ir_3O_8$. Unlike the putative spin-liquid $Na_4Ir_3O_8$, $Na_3Ir_3O_8$ was found to be a semi-metal produced by spin-orbit coupling. The semi-metallic $Na_3Ir_3O_8$ could be an intriguing platform to test the possible non-trivial topological effects, associated with the frustrated and chiral geometry of the lattice and the negative gap produced by SOC. The presence of chirality is particularly unique to the hyper-kagome iridate as compared with the other complex iridium oxides. If single crystals with a sizable size of domains could be grown, we might have a chance to capture topological effects related to the chirality.

**Method**

**Crystal growth and characterizations.** Single crystals of $Na_3Ir_3O_8$ were grown by a flux-growth technique. The mixture of $Na_2CO_3$, $IrO_2$ and $NaCl$ with a ratio of 5:1:20 was loaded in an alumina crucible, and heated up to 1075 °C in an oxygen atmosphere. It was then slowly cooled to 1000 °C with a cooling rate of 2 °C/hour, and subsequently furnace-cooled down to room temperature. Black and block-shaped single crystals were found in the solidified melt. The polycrystalline samples of $Na_4Ir_3O_8$ were synthesized by a conventional solid state reaction as a reference material. The crystal structure of single crystals was analyzed by X-ray diffraction using a three circle diffractometer (Bruker AXS) equipped with SMART APEX II CCD, and Mo $K\alpha$ radiation (see Supplementary). Transport, magnetic and thermodynamic properties were measured with Quantum Design PPMS and MPMS.

In the obtained crystals, we in fact found a very small amount of insulating crystals besides the majority of metallic ones. The X-ray diffraction indicated that the insulating crystals also had the $Na_3Ir_3O_8$ stoichiometry, and the spectroscopic measurements showed no difference between the metallic and insulating crystals. Judging from the poor quality of the insulating crystals, we suspect that the insulating crystals include domains of the $Na_4Ir_3O_8$ phase and there might be a temperature window where $Na_4Ir_3O_8$ phase is stabilized during the crystal growth. However, the growth of $Na_4Ir_3O_8$ single crystal is a future perspective and out of the scope of this article.

**Resonant inelastic x-ray scattering (RIXS).** Resonant inelastic X-ray scattering (RIXS) measurement was performed at BL11XU at SPring-8. Incident X-rays were monochromatized by a double-crystal Si(111) monochromator and a secondary Si(844) channel-cut monochromator. Horizontally scattered X-rays were analyzed by a diced and spherically-bent Si(844) crystal. The total energy resolution was 70 meV. The energy of the incident X-ray was tuned at 11.214 keV, which corresponds to the $L_3$-edge of Ir.

The samples used were single crystal $Na_3Ir_3O_8$ and polycrystalline $Na_4Ir_3O_8$. The spectra were collected at 20 K in $Na_3Ir_3O_8$, whereas at 300 K in $Na_4Ir_3O_8$. The difference in the measurement temperature should not change the overall spectra. The contribution of elastic scattering was evaluated by independently measuring the elastic signal with a $\sigma$-polarized configuration. It was almost negligible in $Na_3Ir_3O_8$, whereas the strong quasi-elastic signal was seen around zero energy in the $Na_4Ir_3O_8$ polycrystalline sample. This is possibly attributed to the elastic signal coming from concomitant Bragg reflections and/or to the phonon excitations inherent to the measurement at room temperature.

**Acknowledgement**

We thank A. W. Rost and A. Kato for technical supports in the RIXS measurement and D. Casa for fabrication of Si(844) analyzer for RIXS. We are grateful to R. Perry, G. Khaliullin, G. Jackeli, D. Pröpper, A. V. Boris and B. Keimer for fruitful discussion. This work was partly supported by Grant-in-Aid for Scientific Research (S) (Grand No. 24224010).

**Author contributions**

T.T. and H.T. conceived and designed the project. T.T., A.M. and J.N. performed synthesis, structural analysis and transport measurements. K.I., M.Y., T.T., A.M. and J.M. measured and analyzed the RIXS data. A.Y. conducted electronic structure calculation. T.T., A.Y. and H.T. wrote the manuscript and all authors discussed and reviewed the paper.

**Competing financial interests**: The authors declare no competing financial interests.

**Figure Captions**

Figure 1 **Crystal structure of $Na_3Ir_3O_8$.** (**a**) The unit cell of $Na_3Ir_3O_8$ with a space group of $P4_132$. The orange, yellow, silver and blue spheres represent Na(1), Na(2), Ir and O atoms, respectively[19]. While Na(1) atoms are enclosed in distorted $O_6$ octahedra, Na(2) atoms are coordinated tetrahedrally by oxygen atoms. The iridium atoms comprise corner-sharing triangles in three dimensions called hyper-kagome lattice. (**b**) Hyper-kagome network of two different chiralities. The red allows represent the orientations of the outer triangles surrounding the centre one. The two hyper-kagome lattices cannot be overlapped with one another.

Figure 2 **Transport, magnetic and thermodynamic properties of $Na_3Ir_3O_8$.** (**a**) Resistivity, (**b**) Hall coefficient and (**c**) Magnetic susceptibility of the single crystalline $Na_3Ir_3O_8$. The red dots in (**c**) represent magnetic susceptibility of $Na_4Ir_3O_8$ polycrystalline sample for comparison. The inset in (**a**) shows specific heat $C$ divided by temperature $T$ of $Na_3Ir_3O_8$. The solid line indicates a fitting line based on the conventional model $C/T = \gamma + \beta T^2$.

Figure 3 **RIXS spectra of $Na_3Ir_3O_8$ single crystal and polycrystalline $Na_4Ir_3O_8$ at Ir $L_3$-edge.** The two spectra were normalized by the intensity of the high energy tail above 6 eV. The blue and pink solid lines around zero energy show the normalized elastic signals independently measured with $\sigma$-polarized incident X-ray.

Figure 4 **Electronic structure calculation for $Na_3Ir_3O_8$.** (a) Scalar-relativistic band structure, and

(b) orbital resolved Ir 5*d* density of states. The *xy* orbital represents a $t_{2g}$ orbital lying in the plane which contains two shorter Ir-O(2) bonds. (c) Relativistic band structure. The bands which form hole and electron pockets are colored in red and magenta, respectively. (d) Ir 5*d* density of states corresponding to (c) resolved with *j* = 5/2 and 3/2 characters.

Table 1 **Refined structural parameters of $Na_3Ir_3O_8$.** The space group is $P4_132$ (No. 213) and $Z = 4$, and the lattice constant is $a = 8.9857(4)$ Å. $g$ and $U_{iso}$ denote site occupancy and isotropic displacement parameter, respectively. The final $R$ indices are $R = 0.0133$ and $wR = 0.0287$.

| Atom | site | g | x | y | z | $U_{iso}$ (Å$^2$) |
|---|---|---|---|---|---|---|
| Ir | 12d | 1 | 0.61264(1) | x + 1/4 | 5/8 | 0.00802(4) |
| Na(1) | 4b | 1 | 7/8 | 7/8 | 7/8 | 0.0122(5) |
| Na(2) | 8c | 1 | 0.2570(2) | x | x | 0.0138(4) |
| O(1) | 8c | 1 | 0.1144(2) | x | x | 0.0105(6) |
| O(2) | 24e | 1 | 0.1364(3) | 0.9071(2) | 0.9186(2) | 0.0111(4) |

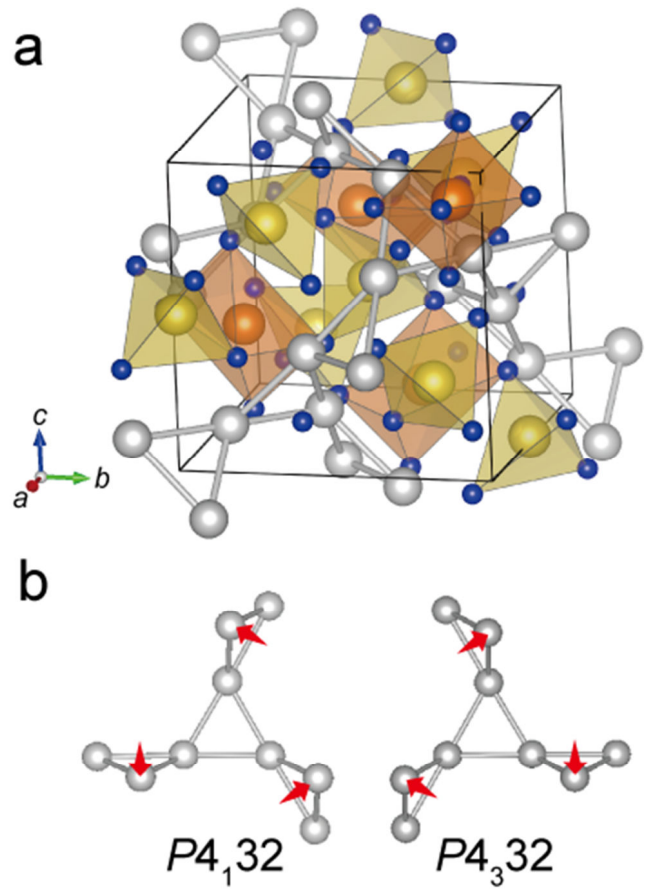

Figure-1 (T. Takayama et al.)

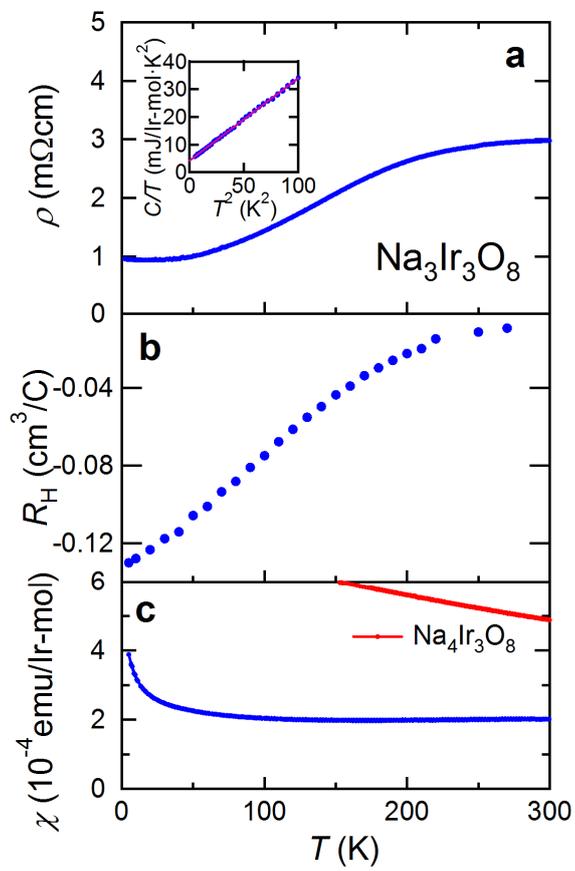

Figure-2 (T. Takayama et al.)

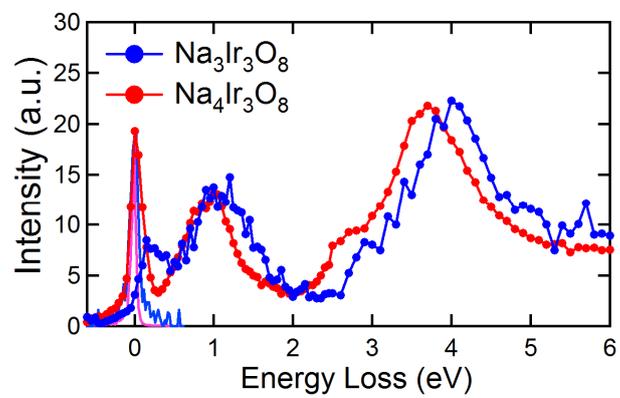

Figure-3 (T. Takayama et al.)

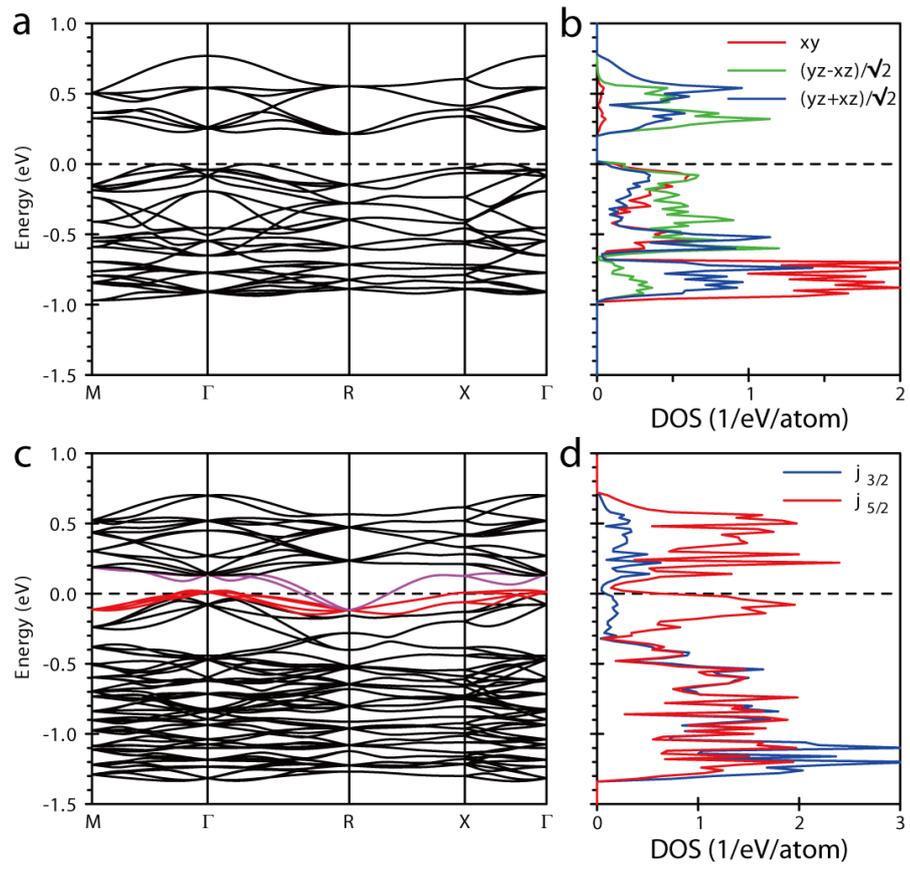

Figure-4 (T. Takayama et al.)